\documentclass[twocolumn,aps,prl]{revtex4} %superscriptaddress,eqsecnum,

\usepackage{amsmath,amssymb,amsfonts,bm}
\usepackage{graphicx,epstopdf}  %epsfig,
\usepackage{color}
\usepackage{lscape}

\usepackage{extarrows}
\usepackage{enumitem}
\usepackage{empheq}

\renewcommand{\d}{{\rm d}}

\newcommand{\w}{\omega}

\newcommand{\B}{\mbox{\tiny B}}

\newcommand{\tS}{\mbox{\tiny S}}
\newcommand{\T}{\mbox{\tiny T}}

\newcommand{\SB}{\mbox{\tiny SB}}

\newcommand{\la}{\langle}
\newcommand{\ra}{\rangle}

\newcommand{\nl}{\nonumber \\}
\newcommand{\be}{\begin{equation}}
\newcommand{\ee}{\end{equation}}
\newcommand{\bsube}{\begin{subequations}}
\newcommand{\esube}{\end{subequations}}
\newcommand{\Eq}[1]{Eq.\,(\ref{#1})}
\newcommand{\Eqs}[1]{Eqs.\,(\ref{#1})}
\newcommand{\Fig}[1]{Fig.\,\ref{#1}}

\newcommand{\RN}[1]{%
  \textup{\uppercase\expandafter{\romannumeral#1}}%
}

\allowdisplaybreaks[1]

\begin{document}

\title{Temperature--dependence of the subdivision potential in nanothermodynamics
}
%%%
\author{Yu Su}
\thanks{Authors of equal contributions}
\author{Zi-Fan Zhu} 
\thanks{Authors of equal contributions}
%\author{Zi-Hao Chen}
%\thanks{Authors of equal contributions}
\author{Hong Gong}
\author{Yao Wang}
\email{wy2010@ustc.edu.cn}
\author{Rui-Xue Xu}
\author{YiJing Yan}
\affiliation{Hefei National Laboratory for Physical Sciences at the Microscale and Department of Chemical Physics, University of Science and Technology of China, Hefei, Anhui 230026, China}

\date{\today}

\begin{abstract}
Nanothermodynamics is the thermodynamics of small systems, which are significantly affected by their surrounding environments. In nanothermodynamics, Hill introduced the concept of subdivision potential, which charaterizes the non-extensiveness. In this work, we establish the quantum thermodynamic integration of the subdivision potential, which is identified to be proportional to the difference between the thermal and von Neumann entropies, focusing on its temperature--dependence. As a result, it serves as a versatile tool to help analyze the origin of non-extensiveness in nanosystems.
\end{abstract}
\maketitle

\paragraph{Introduction --}Systems are inevitably interacting with
the environments in which they are embedded. A system is referred to be small, if it is significantly affected by the surrounding environments. For small systems, classical thermodynamics, remarked by Einstein as ``\emph{the only
physical theory which I am convinced will never be overthrown
within the framework of applicability of its basic concepts}'' \cite{Ein79}, is not directly suitable since the non-extensiveness. In 1962, Hill proposed the nanothermodynamics \cite{Hil623182,Hil63}, which adds an \emph{subdivision potential} term to the traditional Gibbs equation, reading
\begin{align}\label{GH}
E=U+\varepsilon\equiv TS-pV+\mu N+\varepsilon.
\end{align}
Equation (\ref{GH}) is known as  the Gibbs--Hill equation, where $\varepsilon$ is nothing but the subdivision potential.
This approach intrigues widespread interests among various fields of modern sciences \cite{Hil01111,Hil01273,Cha1552,Qia12201,Bed20}.

For a system at finite temperate $T$,  with constant volume $V$ and particle number $N$, the Helmholtz free energy $F=E-TS$ is related to partition function $Z$ as
\be \label{F}
F=-\beta^{-1}\ln Z.
\ee
Throughout this paper, we set $\beta=1/(k_{B}T)$,
with $k_B$ and $T$ being the Boltzmann constant and temperature, respectively.
For simplicity, we always set $k_{B}=1$ below.

For large systems, which is extensive due to the thermodynamic limit, $Z$ characterizes the canonical ensemble, i.e., 
\be \label{Zcanonical}
Z={\rm tr}_{\tS}\big(e^{-\beta H_{\tS}}\big) \longrightarrow\ \ {{\rm cf.\, \Eq{Z}}}
\ee
with $H_{\tS}$ being the system Hamiltonian. This will lead to $\varepsilon=0$, resulting $E=U$ in \Eq{GH}. However, small systems are not distributed canonically due to their interactions with the environments, and in these cases \Eq{Zcanonical} is violated and $\varepsilon\neq 0$. It is therefore necessary to modify the $Z$ in \Eq{Zcanonical} and distinguish between $E$ and $U$.

\paragraph{Subdivision potential proportional to the difference between the thermal and von Neumann entropies --}Now turn to the discussion on the subdivision potential $\varepsilon$ of small systems. Since the system is small, as explained above, the environment and the system--environment interaction are necessarily involved. 
Consider a system--plus--environment composite Hamiltonian reading 
\be \label{Ht}
H_{\T}=H_{\tS}+H_{\SB}+h_{\B},
\ee
where $H_{\tS}$, $H_{\SB}$ and $h_{\B}$ are the system, the system--environment interaction and the environment Hamiltonians, respectively. 
To identify the subdivsion potential $\varepsilon$, one may follow Landsberg and introduce the temperature--dependent Hamiltonian $H_{\tS}(\beta)$ as \cite{Elc57161,Mig202471}
\be \label{Hbeta}
H_{\tS}(\beta) \equiv -\beta^{-1}\big[\ln {\rm tr}_{\B}(e^{-\beta H_{\T}})-\ln Z_{\B}\big],
\ee
with $Z_{\B}={\rm tr}_{\B}(e^{-\beta h_{\B}})$. In this formalism, the system density operator is
\be 
\rho_{\tS}(\beta)=\frac{e^{-\beta H_{\tS}(\beta)}}{Z}=\frac{{\rm tr}_{\B}\big(e^{-\beta H_{\T}}\big)}{Z_{\T}},
\ee
where the partition functions  are [{\rm cf}.\,\Eq{Zcanonical}]
\be \label{Z}
Z={\rm tr}_{\tS}\big[e^{-\beta H_{\tS}(\beta)}\big]=\frac{Z_{\T}}{Z_{\B}}\ \ \text{and}\ \  Z_{\T}={\rm Tr}\big(e^{-\beta H_{\T}}\big).
\ee
In this paper, ${\rm Tr}$, ${\rm tr}_{\tS}$ and ${\rm tr}_{\B}$ represent the total trace, the system and the environmental subspace partial trace, respectively.
Since the bath is less affected by the system, we identify the $E$ as difference between the energy of total system $E_{\T}$ and that of bath $E_{\B}$, leading to the equality 
\be \label{Edef}
E=E_{\T}-E_{\B}=-\frac{\partial \ln (Z_{\T}/Z_{\B})}{\partial \beta}=-\frac{\partial \ln Z}{\partial \beta},
\ee
where $E_{\T} = -\partial\ln Z_{\T}/\partial\beta$ and $E_{\B} = -\partial\ln Z_{\B}/\partial\beta$. On the other hand, the thermodynamic internal energy of system $U$ is considered as the average of $H_{\tS}(\beta)$, namely
\be \label{Udef}
U=\la H_{\tS}(\beta) \ra\equiv Z^{-1}{\rm tr}_{\tS}\big[H_{\tS}(\beta)e^{-\beta H_{\tS}(\beta)}\big].
\ee
%
%Therefore, the subdivision potential reads
%\be \label{eq10}
%\varepsilon=-\frac{\partial \ln Z}{\partial \beta}-\la H_{\tS}(\beta) \ra=\beta\Big\la\frac{\partial H_{\tS}(\beta)}{\partial \beta} \Big\ra,
%\ee
%where we have used \Eq{GH}.
Therefore, we can identify the subdivision potential as [cf.\, \Eqs{GH}, (\ref{Edef}) and (\ref{Udef})]
\be \label{eq10}
\varepsilon
%=\beta\Big\la\frac{\partial H_{\beta}}{\partial \beta} \Big\ra
=-\frac{\partial \ln Z}{\partial \beta}-\la H_{\tS}(\beta) \ra=\beta\Big\la\frac{\partial H_{\tS}(\beta)}{\partial \beta} \Big\ra.
\ee

To exhibit the temperature dependence of $\varepsilon$, we first rewrite $\la H_{\tS}(\beta) \ra$ as
\be 
\la H_{\tS}(\beta) \ra=-\beta^{-1}\big\{{\rm tr}_{\tS}[\rho_{\tS}(\beta)\ln \rho_{\tS}(\beta)]+\ln Z\big\},
\ee
and it together with  \Eq{eq10} shows that
\be\label{key} 
\varepsilon=\beta^{-1}[S_{\rm therm}(\beta)-S_{\rm vN}(\beta)]
\ee
with the thermal entropy:
\begin{align} \label{13_1}
S_{\rm therm}(\beta)=-\frac{\partial F}{\partial T}=\beta^2\frac{\partial F}{\partial \beta},
\end{align}
and the von Neumann entropy:
\begin{align}\label{13_2}
S_{\rm vN}(\beta)=-{\rm tr}_{\tS}[\rho_{\tS}(\beta)\ln \rho_{\tS}(\beta)].
\end{align}
Here, the free--energy $F$ is defined in \Eq{F} with partition function $Z$ in \Eq{Z}. Apparently, \Eq{key} tells us that the subdivision potential $\varepsilon$ is proportional to the difference between the thermal entropy $S_{\rm therm}(\beta)$ and the von Neumann entropy $S_{\rm vN}(\beta)$, as defined in \Eqs{13_1} and (\ref{13_2}), respectively. It is reasonable to render the difference arises from that the thermodynamic limit is not satisfied for small systems.

 \paragraph{Thermodynamic integration --}

Now, we apply the thermodynamic integration for the free energy $F$, as done in our previous works \cite{Gon20154111,Gon20214115}.
\begin{figure}
\includegraphics[width=1.0\columnwidth]{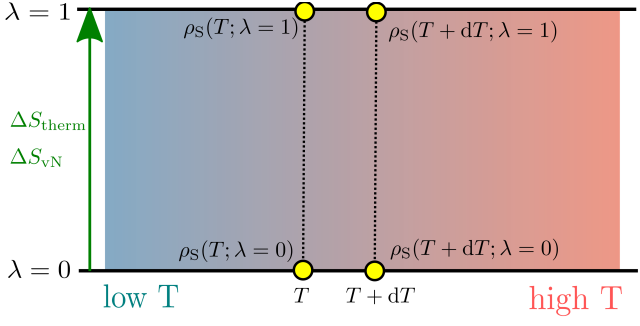}
  \caption{An illustrative depiction of the thermodynamic integration, \Eq{eq15}. The theoretical details can be found in Ref.\,[\onlinecite{Gon20214115}]
} 
\label{fig1}
\end{figure}
Consider a $\lambda$--augmented form of \Eq{Ht}, reading
\be 
H_{\T}(\lambda)=H_{\tS}+\lambda H_{\SB}+h_{\B},
\ee
with $\lambda\in [0,1]$ characterizing the hybridization. Consider the a hybridization process with respect to $\lambda$, at a constant temperature $T$, as depicted in \Fig{fig1}. The free energy change can be written in the integration form \cite{Gon20154111,Gon20214115}
\begin{align}\label{eq15}
\Delta F(T)&\equiv F(T;\lambda=1)-F(T;\lambda=0),
\nl &
=\int_{0}^{1}\!\frac{{\rm d}\lambda}{\lambda}\,{\rm Tr}[(\lambda H_{\SB})\rho_{\T}(T;\lambda)],
\end{align}
with $\rho_{\T}(T;\lambda)=e^{-\beta H_{\T}(\lambda)}/Z_{\T}(\beta;\lambda)$.
 
We can obtain $\Delta F(T)$ in \Eq{eq15} according to the Kirkwood thermodynamics integration as
\begin{align}
\Delta F(T)&= \int^{1}_0\! \frac{\d \Delta F(T)}{\d\lambda}\d\lambda
\nl &
=-\frac{1}{\beta}\int^{1}_0\!\frac{\d\lambda}{Z_{\T}(\lambda)}{\rm Tr}\Big[\frac{\d }{\d\lambda}e^{-\beta(H_{\tS}+h_{\B}+\lambda H_{\SB})}\Big].
\end{align}
It leads to $\Delta F(T)$ in \Eq{eq15}, due to the following relation:
\[
-\frac{1}{\beta}{\rm Tr}\Big[\frac{\d }{\d\lambda}e^{-\beta(H_{\tS}+h_{\B}+\lambda H_{\SB})}\Big]={\rm Tr}\Big[H_{\SB}e^{-\beta(H_{\tS}+h_{\B}+\lambda H_{\SB})}\Big].
\]
This relation is a special case of  the general relation reading
\[
{\rm Tr}\Big[\frac{\d}{\d\lambda}e^{O(\lambda)}\Big]={\rm Tr}\Big[\frac{\d O(\lambda)}{\d\lambda}e^{O(\lambda)}\Big],
\]
which can be proved via the operator  identity
\[
\frac{\d}{\d\lambda}e^{O(\lambda)}=\int_{0}^{1}\!{\d s}\,e^{(1-s)O(\lambda)}\frac{\d O(\lambda)}{\d\lambda}e^{s O(\lambda)}
\]
and the trace cyclic invariance.

As a result, \Eqs{eq10} and (\ref{eq15}) give rise to
\begin{align}\label{eq18}
\varepsilon(T)&=T\left[-\frac{\partial F(T;\lambda=1)}{\partial T}-S_{\rm vN}(\beta)\right]
\nl &
=T\!\left[-\frac{\partial\Delta F(T)}{\partial T}-\frac{\partial F(T;\lambda=0)}{\partial T}-S_{\rm vN}(\beta)\right]
\nl &
=T[\Delta S_{\rm therm}(T)-\Delta S_{\rm vN}(T)].
\end{align}
In \Eq{eq18}, we have defined 
$
\Delta S_{\rm therm/vN}(T)\equiv S_{\rm therm/vN}(T;\lambda=1)-S_{\rm therm/vN}(T;\lambda=0)
$ and used the equality $S_{\rm vN}(T;\lambda=0)=S_{\rm therm}(T;\lambda=0)$ due to the canonicity in the absence of system--bath interactions. 
Numerically, all these quantities can be computed via the dissipaton-equation-of-motion method ($\lambda$-dynamics formalism or imaginary--time formalism) \cite{Yan14054105,Gon20154111,Gon20214115}, which is a second quantization generalization of the well--known hierarachical equations of motion, serving as a rigid approach to the dynamics of a specific system coupled to the Gaussian environments \cite{Tan89101,Tan906676,Yan04216,Xu05041103}.

\begin{figure}[h]
\includegraphics[width=1.0\columnwidth]{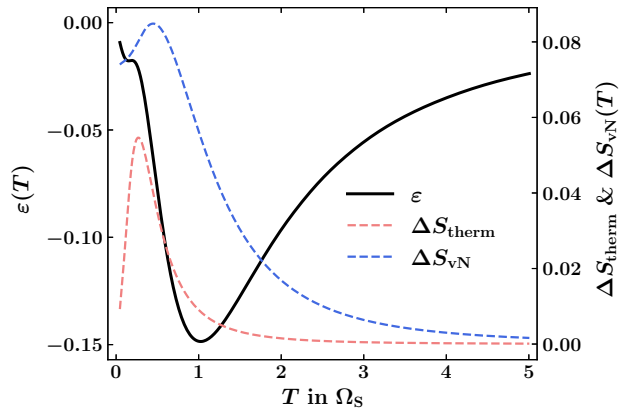}
  \caption{Tempature--dependenece of the subvivision potential. Parameters are taken in accordance to that in Ref.\,[\onlinecite{Gon20154111}]. The bath spectral density $J(\w)=(\pi/2) \sum_j c_j^2\delta (\w-\w_j)$ takes the Drude form as $J(\w)=\eta\gamma\w/(\w^2+\gamma^2)$ with parameters: $\gamma=4V$ and $\eta=0.5V$. The spin energy difference is $E=0.5V$, and the temperature is scaled by the system charateristic frequency $\Omega_{\tS}\equiv 2\sqrt{E^2+V^2}$.
}
\label{fig2}
\end{figure}

\paragraph{Example --} As an example, we consider a spin--boson model 
\be 
H_{\T}(\lambda)=(E\hat{\sigma}_{z}+V\hat{\sigma}_{x})+\lambda \hat{\sigma}_{z} \hat F+h_{\B},
\ee
with $h_{\B}$ being harmonic and
the hybrid bath modes $\hat F$ being linear, i.e.,
$
 h_{\B}=\frac{1}{2}\sum_j\omega_j(\hat p_j^2+\hat x_j^2) \ \ {\rm and}\  \
 \hat F=\sum_j c_{j}\hat x_j.
$
The computational results are shown in \Fig{fig2}, with parameters given in the caption.
In \Fig{fig2}, we show the temperature dependence of $\varepsilon(T)$, as well as $\Delta S_{\rm therm}(T)$ and $\Delta S_{\rm vN}(T)$. The temperature is scaled by the system charateristic frequency $\Omega_{\tS}\equiv 2\sqrt{E^2+V^2}$ \footnote{As mentioned above, we have set $k_{B}=1$.}.
It is observed that the subdivision potential exhibits a turnover near the characteristic frequency, and this is conjectured as a universal feature.

 \paragraph{Summary --}
 In summary, we first identify the subdivision potential $\varepsilon$ to be proportional to the difference between the thermal and von Neumann entropies, followed by the establishment of its quantum thermodynamic integration. We explicitly show its temperature--dependence by taking the spin--boson model as an example, and one characteristic turnover point of subdivision potential is observed. %And it is believed that turnover behavior might be measured in experiment. 
 Besides, the method developed in this work serves as a versatile tool to help analyze the origin of non-extensiveness in nanosystems.

\vspace{1em}

Support from the Ministry of Science and Technology of China, Grant No.\ 2017YFA0204904, and the National Natural Science Foundation of China, Nos. 21633006, 22103073 and 22173088 is gratefully acknowledged. This research is partially motivated by the 1st ``Question and Conjecture'' activity supported by the Top Talent Training Program 2.0 for undergraduates.

%Support from the the National Natural Science Foundation of China, No. 22103073 is gratefully acknowledged.

% \bibliographystyle{./aip}
% \bibliography{./bibrefs}

\end{document}